\documentclass{aa}  

\usepackage{graphicx}
\usepackage{txfonts}
\usepackage{xcolor}
\usepackage{ulem}
\usepackage{hyperref}
\usepackage{orcidlink}
\hypersetup{
  colorlinks,
  citecolor=blue,
  linkcolor=blue,
  urlcolor=blue}

\begin{document} 

   \title{Large-scale ordered magnetic fields generated in mergers of helium white dwarfs}

   \author{R\"udiger Pakmor\inst{1}\thanks{rpakmor@mpa-garching.mpg.de}
           \orcidlink{0000-0003-3308-2420}
           \and
           Ingrid Pelisoli\inst{2}\orcidlink{0000-0003-4615-6556}
           \and
           Stephen Justham\inst{1}\orcidlink{0000-0001-7969-1569}
           \and
           Abinaya S. Rajamuthukumar\inst{1}\orcidlink{0000-0002-1872-0124}
           \and\newline
           Friedrich K. R\"opke\inst{3,4,5}\orcidlink{0000-0002-4460-0097}
           \and
           Fabian R. N. Schneider\inst{3,5}\orcidlink{0000-0002-5965-1022}
           \and
           Selma E. de Mink\inst{1}\orcidlink{0000-0001-9336-2825}
           \and
           Sebastian T. Ohlmann\inst{6}\orcidlink{0000-0002-6999-1725}
           \and\newline
           Philipp Podsiadlowski\inst{7}
           \and
           Javier Mor\'an-Fraile\inst{3} \orcidlink{0000-0002-8918-5130}
           \and
           Marco Vetter\inst{3,5}\orcidlink{0009-0007-2322-6001}
           \and
           Robert Andrassy\inst{5,3}
          }

   \institute{Max-Planck-Institut f\"{u}r Astrophysik, 
              Karl-Schwarzschild-Str. 1, D-85748, Garching, Germany
         \and
              Department of Physics, University of Warwick, Gibbet Hill Road, Coventry, CV4 7AL, UK
         \and
              Heidelberger Institut f{\"u}r Theoretische Studien, Schloss-Wolfsbrunnenweg 35, 69118 Heidelberg, Germany
         \and
              Zentrum f{\"u}r Astronomie der Universit{\"a}t Heidelberg, Institut f{\"u}r Theoretische Astrophysik, Philosophenweg 12, 69120 Heidelberg, Germany
         \and
              Zentrum f{\"u}r Astronomie der Universit{\"a}t Heidelberg, Astronomisches Rechen-Institut, M{\"o}nchhofstr, 12-14, 69120 Heidelberg, Germany
         \and
              Max Planck Computing and Data Facility, Gießenbachstraße 2, 85748 Garching, Germany
         \and
              University of Oxford, St Edmund Hall, Oxford OX1 4AR, UK
             }

   \date{}
 
  \abstract
   {
   Stellar mergers are one important path to highly magnetised stars. Mergers of two low-mass white dwarfs may create up to every third hot subdwarf star. The merging process is usually assumed to dramatically amplify magnetic fields. However, so far only four highly magnetised hot subdwarf stars have been found, suggesting a fraction of less than $1\%$.

   We present two high-resolution magnetohydrodynamical (MHD) simulations of the merger of two helium white dwarfs in a binary system with the same total mass of $0.6\,M_\odot$. We analysed an equal-mass merger with two $0.3\,M_\odot$ white dwarfs, and an unequal-mass merger with white dwarfs of $0.25\,M_\odot$ and $0.35\,M_\odot$. We simulated the inspiral, merger, and further evolution of the merger remnant for about $50$ rotations.

   We found efficient magnetic field amplification in both mergers via a small-scale dynamo, reproducing previous results of stellar merger simulations. The magnetic field saturates at a similar strength for both simulations.

   We then identified a second phase of magnetic field amplification in both merger remnants that happens on a timescale of several tens of rotational periods of the merger remnant. This phase generates a large-scale ordered azimuthal field via a large-scale dynamo driven by the magneto-rotational instability.

   Finally, we speculate that in the unequal-mass merger remnant, helium burning will {initially} start in a shell around a cold core, rather than in the centre. This forms a convection zone that coincides with the region that contains most of the magnetic energy, and likely destroys the strong, ordered field. Ohmic resistivity might then quickly erase the remaining small-scale field. Therefore, the mass ratio of the initial merger could be the selecting factor that decides if a merger remnant will stay highly magnetised long after the merger.
   }


   \maketitle

\section{Introduction}

Stellar mergers are thought to result in highly magnetised stars. The amplification of magnetic fields in stellar mergers is challenging to model directly though, because magnetic dynamos are inherently three-dimensional (3D) processes that operate on dynamical timescales. Directly modelling magnetic dynamos therefore requires 3D magneto-hydrodynamical (MHD) simulations.

The first 3D stellar merger simulations that include magnetic fields have recently shown that magnetic fields are essentially always amplified in the Kelvin-Helmholtz unstable shear layers and the turbulence induced there. After the merger, the remnant is  left with a strong unordered magnetic field with many field reversals. This result has consistently been found for systems as different as mergers of carbon-oxygen white dwarfs in binary systems \citep{Zhu2015}, white dwarf merger remnants \citep{Ji2013}, common envelopes \citep{Ohlmann2016, Ondratschek2022, Gagnier2024}, mergers of main sequence stars \citep{Schneider2019,Schneider2020}, neutron star mergers \citep{Kenta2024}, or neutron star white dwarf mergers \citep{MoranFraile2024}. However, how this magnetic field evolves on much longer timescales, and how much of it will be left millions of years after the merger is an unsolved question \citep{Schneider2019,Schneider2020}. In particular, mergers of low-mass carbon-oxygen white dwarfs are one of the main channels invoked to explain magnetised white dwarfs \citep{Zhu2015,Bagnulo2022}.

\begin{figure*}
    \centering
    \includegraphics[width=0.95\textwidth]{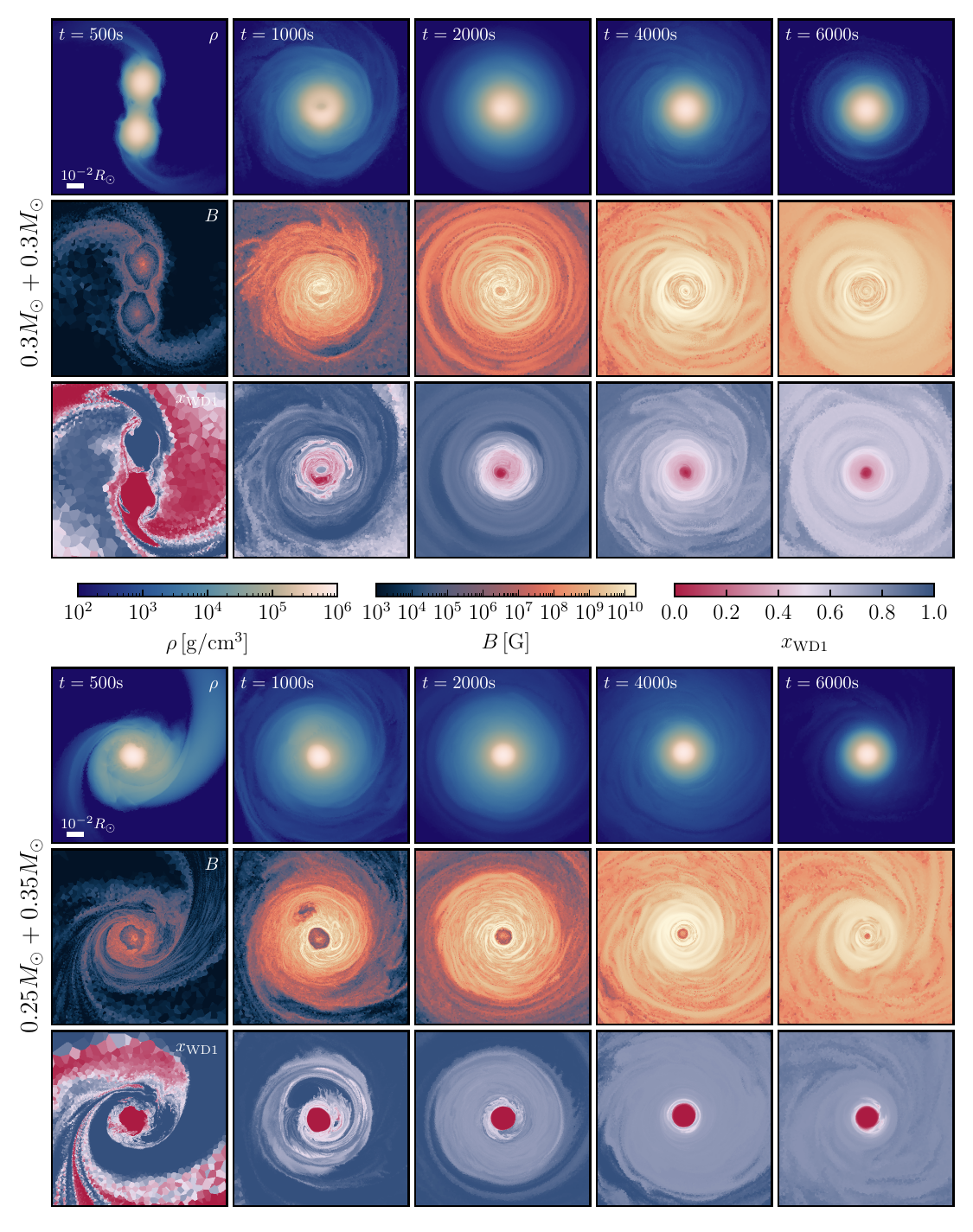}
    \caption{Time evolution of both merger simulations until $6000\,\mathrm{s}$ for the equal-mass merger (top panels) and the unequal mass merger with a mass ratio of $q=0.7$ (bottom panels). The rows show slices of density, magnetic field strength, and mass fraction of the material originally part of one of the white dwarfs in the mid-plane. Both mergers show a large amplification of the magnetic field. In the equal-mass merger, the merger remnant is almost fully mixed. In the unequal-mass merger, the centre of the merger remnant consists of the unmixed core of the initially more massive white dwarf.}
    \label{fig:evolution}
\end{figure*}

Hot subdwarf (sdB/sdO) stars are low-mass stars with a typical mass of about $0.5\,M_\odot$, which have almost no hydrogen left and are powered by helium burning. They are particularly interesting in the context of mergers because they cannot form as single stars in the galaxy, but their formation requires binary interaction \citep{hani,hanii,podsiformation08,hewdsimon,hewdmszhang,clausenwade}. One of the paths to form a hot subdwarf star is the merger of two helium white dwarfs, which themselves require binary interaction to form \citep{webbinkhewd,hewdsimon,justhamco,hewdmszhang}. This channel likely contributes a significant fraction, possibly up to a third of all hot subdwarf stars in a steady-state Galactic population \citep{hanii}, and is plausibly dominant in populations older than ${\gtrsim}10\,\mathrm{Gyr}$ \citep{Han2008}.

If all sdB/sdO stars that are formed via mergers of white dwarfs were highly magnetised, and the fields were long lived, we would expect a large fraction of sdB/sdO stars to be highly magnetised. So far, however, only four highly magnetised sdO stars have been found with surface strengths of several $10^5\,\mathrm{G}$ \citep{Dorsch2022,Pelisoli2022}. Surprisingly, those also share their spectral sub-type and have similar positions in the Hertzsprung–Russell diagram. Not only are they all helium sdOs, but they are all of an intermediate-helium rich sub-type which is particularly unusual at their high effective temperature \citep{Dorsch2022}. This could be an indication that not all hot subdwarf stars that formed from white dwarf mergers become and remain highly magnetised, but that other special conditions are required.

Here we present two 3D MHD merger simulations of binary systems consisting of two helium white dwarfs with the same total mass but a different mass ratio. We followed them through inspiral and merger and evolved the merger remnants for more than $50$ rotations. We focus on the amplification and structural evolution of the magnetic fields. We characterise and discuss the similarities and differences between the magnetic field configuration and the likely long-term evolution of the merger remnants in the equal-mass merger and the unequal-mass merger.

The paper is structured as follows. In Section~\ref{sec:methods} we describe the methods and the setup of our simulations. In Section~\ref{sec:merger} we summarise the inspiral and merger phase and the magnetic field amplification during this period. In Section~\ref{sec:remnant} we then analyse the longer-term evolution of the merger remnants with a focus on the origin of the large-scale dynamo. We discuss the possible further evolution of both merger remnants far beyond the timescales we can directly simulate and broader implications of our results in Section~\ref{sec:discussion} and finish with a summary and outlook in Section~\ref{sec:summary}.

\section{Simulations}
\label{sec:methods}

We simulate the merger of two helium white dwarfs with the moving-mesh code \textsc{arepo} \citep{Arepo, Pakmor2016, Weinberger2020}. It solves ideal MHD on an unstructured Voronoi mesh using a second-order finite-volume scheme \citep{Pakmor2016}. \textsc{arepo} moves a set of mesh-generating points that create the Voronoi mesh with the gas flow to obtain an almost Lagrangian scheme that maintains the accuracy of traditional finite volume schemes.

The ideal MHD solver of \textsc{arepo} \citep{Pakmor2011b, Pakmor2013b} uses the Powell scheme for divergence control \citep{Powell1999}. \textsc{arepo} solves for self-gravity using a hierarchical oct-tree, that is fully coupled to the MHD. Finally, we employ the Helmholtz equation of state that includes ions as non-relativistic ideal gas with Coulomb corrections, an arbitrarily degenerate electron-positron gas, and radiation \citep{Timmes2000}. We ignore nuclear reactions, because the temperatures do not significantly exceed $10^8\,\mathrm{K}$ and the energy release from nuclear reaction on the timescales we simulate therefore remains negligible.

We start by generating one-dimensional profiles of individual pure helium white dwarfs with constant temperature $5\times 10^5\,\mathrm{K}$ in hydrostatic equilibrium. We neglect any hydrogen envelope often present on the surface of helium white dwarfs. We note that the choice of this initial temperature is irrelevant for our simulations, because the thermal energy is dynamically unimportant and the pressure is completely dominated by the degeneracy pressure of electrons. The profiles are therefore a one parameter family and they are fully determined by the total mass of the helium white dwarf.

We generate a 3D Voronoi mesh in \textsc{arepo} with cells of roughly equal mass of $10^{-7}\,M_\odot$ \citep{Pakmor2012}. We use the density and pressure profile to initialise the properties of the cells. We then relax each white dwarf in isolation for five dynamical timescales with a damping term that decreases with time \citep{Ohlmann2017}. After that, we evolve each white dwarf further in isolation for another five dynamical timescales to make sure that all white dwarfs are fully stable in hydrostatic equilibrium.

We add a dipole magnetic field to the relaxed white dwarfs with a strength at the surface of $10^3\,\mathrm{G}$. At this strength the magnetic field is dynamically completely irrelevant, that is the magnetic pressure is many orders of magnitude smaller than the total gas pressure. In the centre, we soften the radial dependence of the dipole magnetic field strength with a radius of $10^{-3}\,\mathrm{R_\odot}$. We note that we simulate ideal MHD, that is we neither include explicit viscosity nor resistivity. Therefore any kinetic viscosity and magnetic resistivity is purely numerical. The resulting effective magnetic Prandtl number, that is the ratio between the magnetic and kinetic Reynolds number resulting from the numerical viscosity and resistivity, is likely of order unity, because the dissipation scales for magnetic and velocity fields are both the grid scale. We discuss estimates and implications of the physical Prandtl number of our systems in the different phases of the merger in Section~\ref{sec:discussion}.

We create two binary systems made of the same total mass of $0.6\,M_\odot$. One of the binaries is made of two equal-mass helium white dwarfs of $0.3\,M_\odot$ and the other binary is made of two helium white dwarfs of $0.25\,M_\odot$ and $0.35\,M_\odot$, respectively. The $0.3\,M_\odot$ white dwarfs have a central density of $5.2\times 10^{5}\,\mathrm{g\,cm^{-3}}$. The other white dwarfs have central densities of $7.5\times 10^{5}\,\mathrm{g\,cm^{-3}}$ for the $0.35\,M_\odot$ white dwarf and $3.4\times 10^{5}\,\mathrm{g\,cm^{-3}}$ for the $0.25\,M_\odot$ white dwarf. Their ratio of central densities is $q_\mathrm{c}=0.45$. This places the binary system firmly into the regime of unequal-mass mergers that is characterised by $q_\mathrm{c}<0.6$ \citep{Zhu2013}.

We create both binary systems on a circular co-rotating orbit with an initial orbital separation of $0.1\,R_\odot$ corresponding to an initial orbital period of $400\,\mathrm{s}$. We put the binary system at the centre of a box with a side length of $10^{12}\,\mathrm{cm}$. This box is large enough that no information travels from the binary at the centre to the edges of the box until the end of our simulation at $6000\,\mathrm{s}$. The white dwarfs are modelled by a total of $6$ million cells. We use explicit refinement and de-refinement during the simulation to ensure that the mass of the cells are within a factor of two of our target gas mass of $10^{-7}\,M_\odot$. We fill the box in the background with helium with a density of $10^{-5}\,\mathrm{g\,cm^{-3}}$, because \textsc{arepo} cannot deal with a true vacuum. Despite being about $20$ orders of magnitude larger than typical densities of the interstellar medium, this adds only $5{\times}10^{-3}\,M_\odot$ of mass and a similarly negligible amount of energy to our simulations. It therefore does not affect any relevant aspects of our results.

We simulate the inspiral of the binary systems for an initial $360\,\mathrm{s}$ with an effectively accelerated gravitational wave emission-like term that decreases the separation of the binary system at a constant rate of $\dot{a}=100\,\mathrm{km\,s^{-1}}$ \citep{Pakmor2021,Pakmor2022} to obtain a close relaxed binary system quickly, but slowly enough so that both stars remain in equilibrium. After $360\,\mathrm{s}$ we continue both simulations until $6000\,\mathrm{s}$ fully self-consistently without this extra term, and angular momentum and total energy are conserved from this time onwards.

Both mergers are conservative for the time we simulate, that is essentially no mass becomes unbound and the bound merger remnant contains the total mass of both original white dwarfs. At the time we stop the inspiral, both binaries have an orbital period of $140\,\mathrm{s}$ and a very similar total angular momentum of $1.59{\times}10^{50}\,\mathrm{g\,cm^2\,s^{-1}}$ for the equal-mass merger and $1.55{\times} 10^{50}\,\mathrm{g\,cm^2\,s^{-1}}$ for the unequal-mass merger. Almost all of the angular momentum is orbital angular momentum. The spins of both white dwarfs contribute less than $5\%$.

\section{Disruption and merger}
\label{sec:merger}

\begin{figure}
    \centering
    \includegraphics[width=\linewidth]{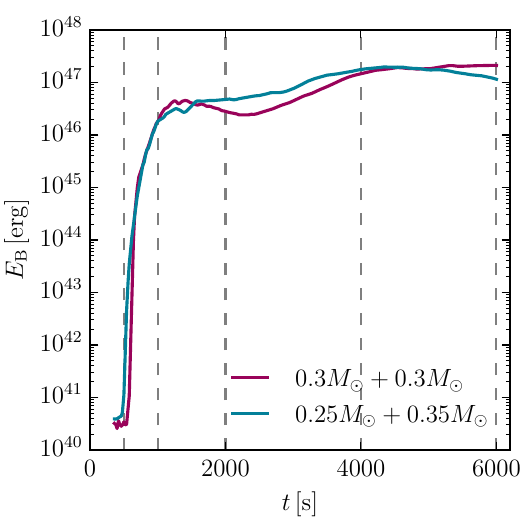}
    \caption{Time evolution of the total magnetic energy in both merger simulations. Vertical dashed lines show the times of the slices in Figure~\ref{fig:evolution}. In both simulations the magnetic energy is amplified by more than five orders of magnitude in the first $1000\,\mathrm{s}$ in the actual merger. Then there is an additional phase of magnetic field amplification starting around $2000\,\mathrm{s}$.}
    \label{fig:evolution_bfld}
\end{figure}

We show the time evolution of both merger simulations and their merger remnant in Figure~\ref{fig:evolution}. Its rows show slices of density, magnetic field strength, and the mass fraction of material that originates from the white dwarf that has a larger mass fraction in the centre at the end of the simulations. For the unequal mass merger, this is the more massive white dwarf.

In the equal-mass merger (shown in the upper half of Figure~\ref{fig:evolution}), both white dwarfs are simultaneously disrupted and first form a low-density cavity in the centre. They then merge into a single object with a central density peak. The magnetic field is quickly amplified first in shear layers, then everywhere in the merger remnant. The merger remnant is almost fully mixed and it only shows a slight preference for material of one of the white dwarfs in the centre, as a result of symmetry breaking by round-off errors that are already present in the initial setup and that are continuously added during the evolution of the simulation. We quantify the mixing at the end of the simulation in Section~\ref{sec:remnant}.

In the unequal-mass merger (shown in the lower half of Figure~\ref{fig:evolution}) the secondary, the less massive white dwarf, is disrupted and mixes with the outer layers of the primary white dwarf. However, the core of the primary white dwarf, consisting of ${\sim}\,0.2\,M_\odot$, remains unaffected. The magnetic field is amplified also in shear layers first, then everywhere in the mixed envelope. The magnetic field in the inert core is not amplified, but numerical diffusion transports the strong magnetic field from the envelope inwards with time. The core of the merger remnant is purely made from material from the core of the primary white dwarf. The envelope is fully mixed and dominated by material from the secondary white dwarf. The diffusion of the magnetic field into the core comes without mixing of material, a strong sign of the numerical nature of this process.

We show the evolution of the total magnetic energy in the simulation for both mergers in Figure~\ref{fig:evolution_bfld}. Both mergers exhibit the same initial fast increase of magnetic energy during the merger by about six orders of magnitude. They then initially saturate around the same value at $1000\,\mathrm{s}$. We first focus on this part and discuss the later evolution of the merger remnants in Section~\ref{sec:remnant}. The initial phase is very similar to the magnetic field amplification seen in simulations of mergers of main sequence stars \citep{Schneider2019, Schneider2020}, common envelope systems \citep{Ohlmann2016,Ondratschek2022}, white dwarf--neutron star mergers \citep{MoranFraile2024}, or mergers of two neutron stars \citep{Kenta2024}. It also erases any dependence on the initial seed field.

More quantitatively, we can try to roughly estimate the physically expected amplification rate of the magnetic field. Assuming Braginskii viscosity \citep{Spitzer1962,Zuhone2015},
\begin{equation}
    \nu \approx 2\times 10^{-16} \left( \frac{T}{\mathrm{K}} \right)^{5/2} \left( \frac{\rho}{\mathrm{g\,cm^{-3}}} \right)^{-1} \,\mathrm{cm^2\,s^{-1}},
\end{equation}
and plugging in values for the shear layer between both white dwarfs, that is a temperature of $T{\sim}5{\times}\,10^7\,\mathrm{K}$ and a density of $\rho{\sim}10^2\,\mathrm{g\,cm^{-3}}$ we obtain a physical viscosity of $\nu{\sim}40\,\mathrm{cm^2\,s^{-1}}$.

If we further assume that the binaries just prior to the full merger drive turbulence in the shear layer between the white dwarfs, we can approximate the outer velocity scale with the relative velocity between both white dwarfs $u{\sim}10^8\,\mathrm{cm\,s^{-1}}$ and the integral scale length of the system with the distance between both white dwarfs $l{\sim}3{\times}10^9\,\mathrm{cm}$. With this we obtain a physical Reynolds number of
\begin{equation}
    \mathrm{Re}\approx\frac{u \cdot l}{2\pi \nu} \sim 10^{15}.
\end{equation}
We can then further estimate the expected amplification rate of the magnetic field in the shear layer for a small scale dynamo in the kinetic regime driven by Kelvin-Helmholtz and Rayleigh–Taylor instabilities following \citet{Skoutnev2021} as
\begin{equation}
    \gamma \approx \frac{u}{2l}\, \mathrm{Re}^{1/2} \sim 5{\times}10^5 \mathrm{s^{-1}}.
\end{equation}
This physical rate is much faster than the amplification rate in our simulations (see Figure~\ref{fig:evolution_bfld}) of $\gamma{\sim}0.04\,\mathrm{s^{-1}}$ for the equal-mass merger and $\gamma{\sim}0.02\,\mathrm{s^{-1}}$ for the unequal-mass merger. This difference is expected because the numerical viscosity in our simulations is much higher than the physical viscosity of the system, and because the turbulent layer is initially far from volume filling. Therefore we expect the dynamo to be slower in our simulations, but still saturate at essentially the same strength \citep[see, e.g.][]{Kriel2023}. Indeed, the dynamo stops and the magnetic energy initially saturates around $1000\,\mathrm{s}$ when the magnetic energy density reaches about $10\%$ of the turbulent energy density, that is the kinetic energy density after subtracting rotation, as expected for a small-scale turbulent dynamo. For an effective Reynolds number of $\mathrm{Re}{\sim}10$, we obtain a $\gamma {\sim}0.05\,\mathrm{s^{-1}}$, similar to the amplification rate we find in the simulations. 

\begin{figure}
    \centering
    \includegraphics[width=\linewidth]{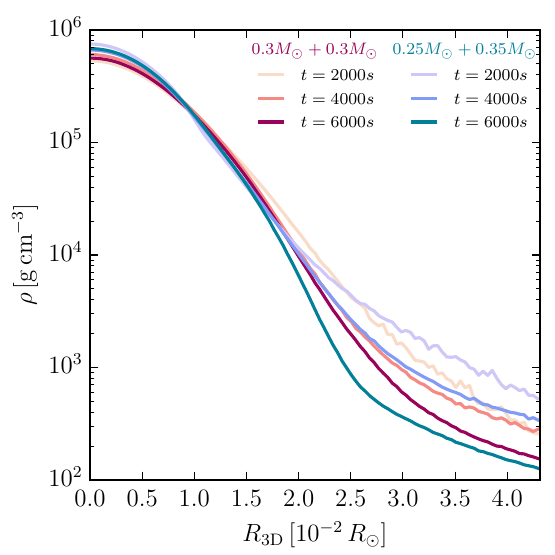}
    \caption{Spherically averaged density profiles of both merger remnants at $2000\,\mathrm{s}$, $4000\,\mathrm{s}$, and $6000\,\mathrm{s}$ (left panel). The density profiles agree between the two mergers for the whole range shown and evolve only slowly.}
    \label{fig:profile_density}
\end{figure}

\begin{figure*}
    \centering
    \includegraphics[width=\textwidth]{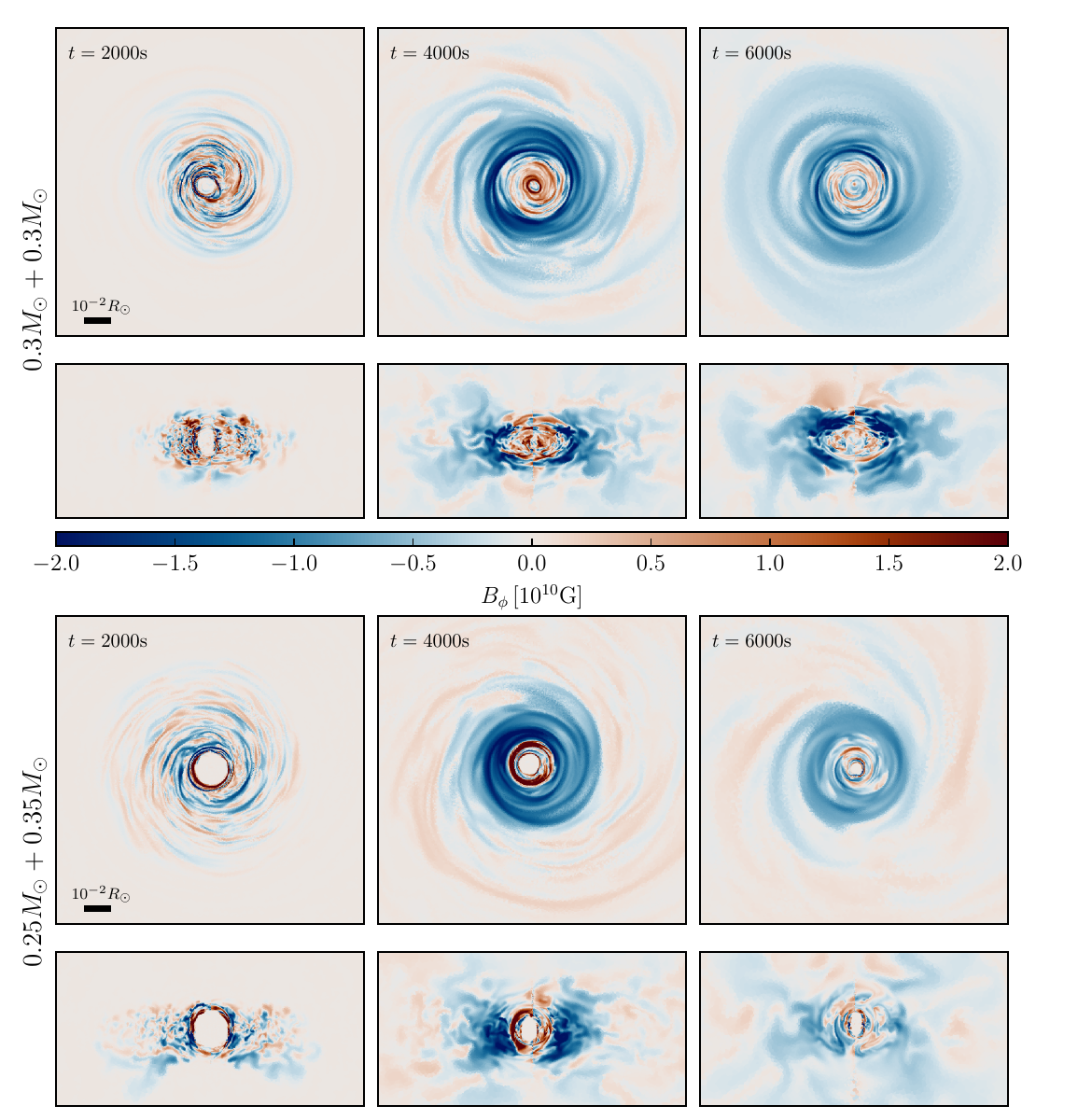}
    \caption{Slices of azimuthal magnetic field strength at $2000\,\mathrm{s}$, $6000\,\mathrm{s}$, and $6000\,\mathrm{s}$ for the equal-mass merger (top half of the figure) and unequal-mass merger (bottom half of the figure). Both merger remnants change from a mostly chaotic small-scale field at $2000\,\mathrm{s}$ to a large-scale ordered azimuthal field at $6000\,\mathrm{s}$. The azimuthal field of the equal-mass merger remnant is slightly stronger and ordered out to larger radii compared to the remnant of the unequal-mass merger.}
    \label{fig:phi}
\end{figure*}

\begin{figure*}
    \centering
    \includegraphics[width=\textwidth]{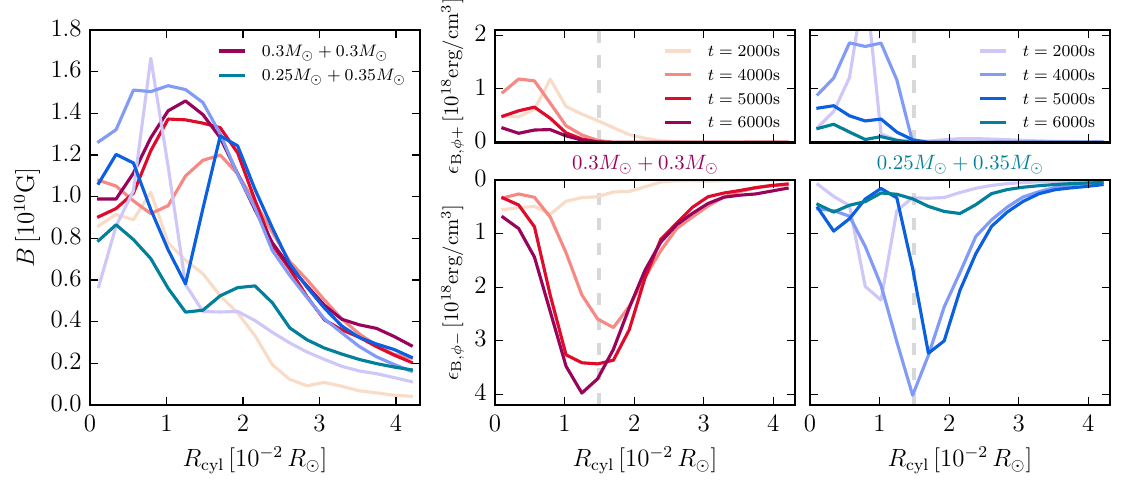}
    \caption{Evolution of the radial profile of the volume weighted root mean square magnetic field strength (left panel) in a cylindrical disk with a height of $0.03\,R_\odot$ centred on the mid-plane. The middle and right panels show the radial profiles of magnetic energy density in the azimuthal component of the magnetic field for the equal mass (red and middle panel) and unequal mass (blue and right panels) mergers. Their upper and lower panels shows the energy of all cells with a positive and negative azimuthal field component, respectively. The vertical line shows the radius of $0.015\,R_\odot$ used in Figure~\ref{fig:butterfly}. Both merger remnants significantly amplify the azimuthal magnetic field that is anti-aligned with the direction of rotation and it completely dominates over the aligned azimuthal field at $4000\,\mathrm{s}$. In the unequal-mass merger remnant the azimuthal magnetic field then decays again significantly until $6000\,\mathrm{s}$. In contrast, it remains strong in the equal-mass merger remnant until the end of the simulation without any sign of decay.}
    \label{fig:profile_bfld}
\end{figure*}

\begin{figure*}
    \centering
    \includegraphics[width=\textwidth]{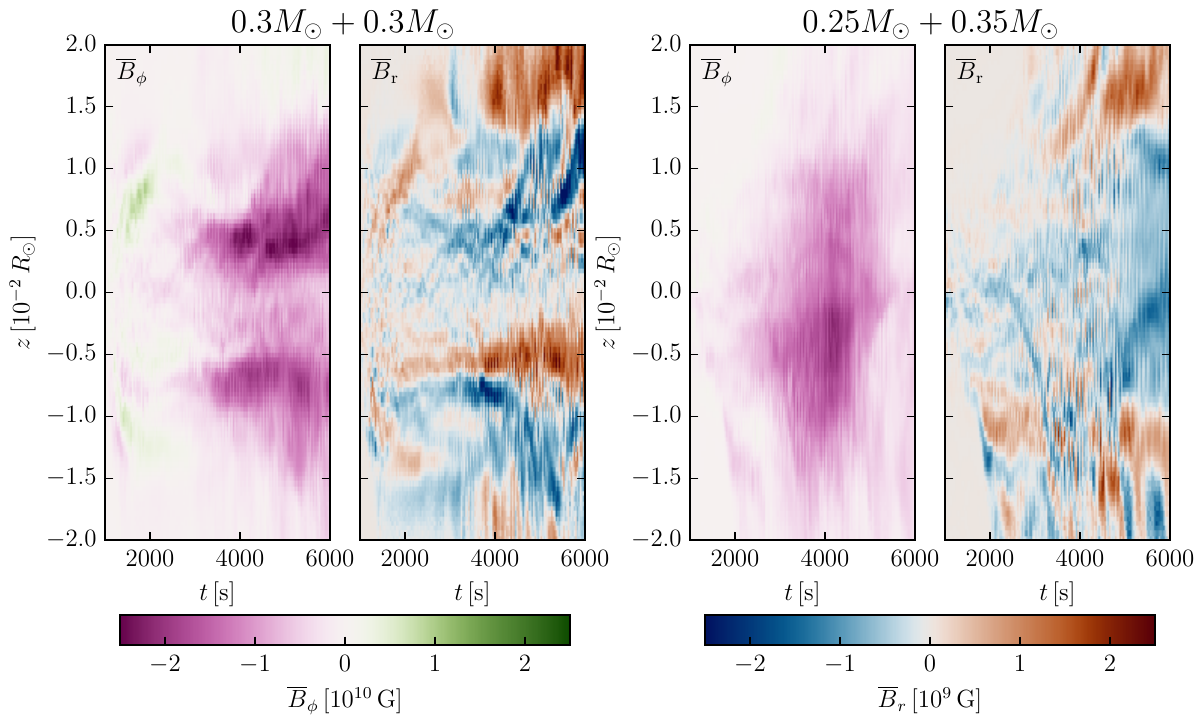}
    \caption{Butterfly diagram at a cylindrical radius of $0.015\,R_\odot$. The left two panels of the plot show the equal-mass merger, the right two panels the unequal-mass merger. The panels show the time evolution of the mean azimuthal (first and third panel) and radial (second and fourth panel) magnetic field versus height. The radial magnetic field shows the outward moving patterns that are characteristic for the MRI. The azimuthal field becomes a large-scale ordered, anti-aligned field.}
    \label{fig:butterfly}
\end{figure*}

\begin{figure*}[h!]
    \centering
    \includegraphics[width=\textwidth]{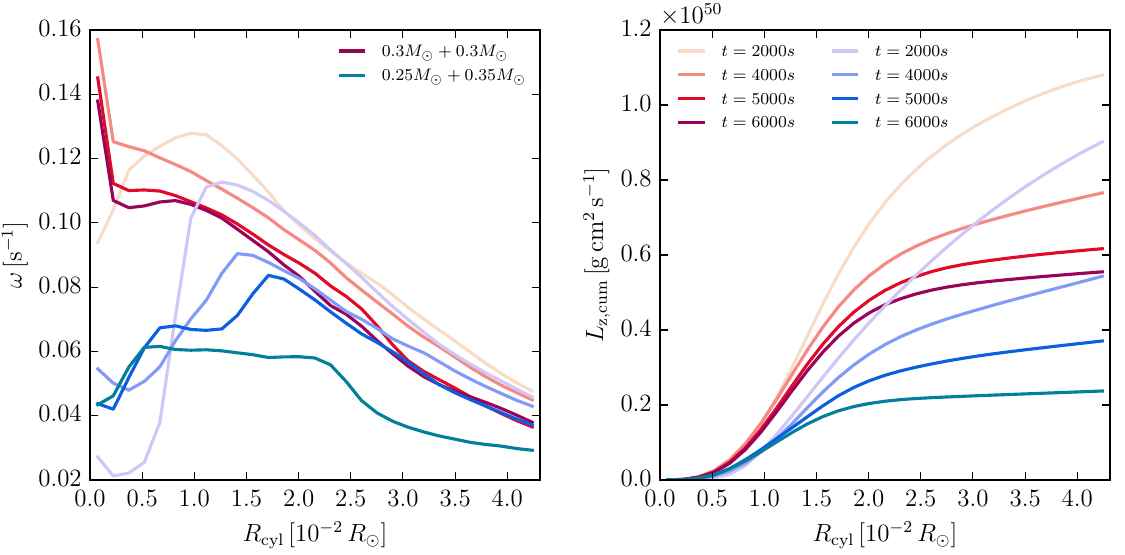}
    \caption{Evolution of the radial profile of the angular velocity in a cylindrical disk with a height of $0.03\,R_\odot$ centred on the mid-plane (left panel) and of the cumulative angular momentum (right panel). The remnant of the unequal-mass merger (blue) initially has a declining omega profile for $R\,>\,0.01\,R_\odot$, but eventually evolves to a flat profile in angular velocity with a local jump around $0.025\,R_\odot$. Declining rotation profiles are unstable to the MRI. The remnant of the equal-mass merger keeps an outward, constantly declining angular velocity profile. The equal-mass merger (red) has and retains significantly more angular momentum in its inner parts shown here.}
    \label{fig:profile_omega}
\end{figure*}

\section{Dynamical evolution of the merger remnant}
\label{sec:remnant}

We now want to understand the evolution of the merger remnants, with a focus on the evolution of their magnetic fields. In particular, we aim to understand the second phase of magnetic field growth seen in Figure~\ref{fig:evolution_bfld} after $2000\,\mathrm{s}$ and the physical processes driving it.

At this time both merger remnants have evolved into differentially rotating objects that are close to spherically symmetric with a central density peak. In both merger remnants the magnetic field is dominated by the unordered magnetic field generated in the small-scale dynamo. This field is many orders of magnitude stronger than the initial seed fields of the individual white dwarfs and independent of their structure and strength. Then a second phase of magnetic field amplification starts that ends around ${\sim}\, 5000\,\mathrm{s}$. In this phase the magnetic energy increases exponentially again, but at a much slower rate than during the initial small-scale dynamo. In the last $1000\,\mathrm{s}$ the magnetic field in the unequal-mass merger remnant decreases again, but the magnetic field in the equal-mass merger remnant remains stable in its saturated state.

We first look at the evolution of the density profile of both merger remnants in Figure~\ref{fig:profile_density}. At $4000\,\mathrm{s}$, both merger remnants have relaxed to essentially the same density profile out at least $0.02\,\mathrm{R_\odot}$ corresponding to a mass coordinate of ${\sim}\,0.5\,M_\odot$. Both density profiles then only evolve slightly until the end of the simulation at $6000\,\mathrm{s}$. Their central density at $6000\,\mathrm{s}$ is still roughly a factor of $4$ lower than the central density of a theoretical cold non-rotating helium white dwarf of the same total mass of $0.6\,M_\odot$ that would have a central density of $3.4{\times}10^6\,\mathrm{g\,cm^{-3}}$. The lower central density of the merger remnants is a result of rotation helping to hold it from immediate contraction. Thermal pressure of the ions is still irrelevant, and the pressure is dominated by the degeneracy pressure of the electrons. The inner envelope at $R\,{\sim}\,0.01\,R_\odot$ rotates with a typical period of $100\,\mathrm{s}$. Thus, at the end of the simulation at $6000\,\mathrm{s}$ about $50$ rotations have passed.

\begin{figure*}
    \centering
    \includegraphics[width=\textwidth]{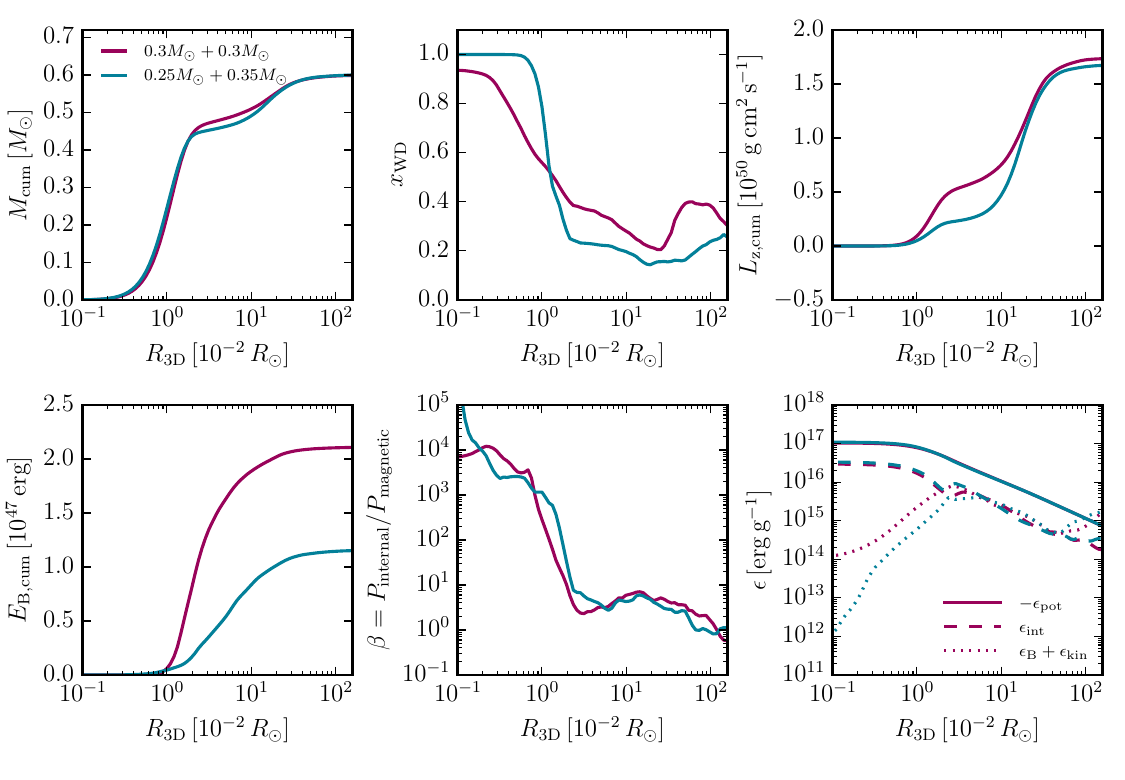}
    \caption{Radial profiles in spherical shells of the merger remnants at the end of the simulations ($6000\,\mathrm{s}$). The panels show cumulative mass (top left), mass fraction of the material that originates from the white dwarf that eventually dominates the centre of the remnant, that is the more massive white dwarf for the unequal-mass merger (top centre),
    cumulative angular momentum (top right), cumulative magnetic energy (bottom left), plasma beta (bottom centre), and specific internal, magnetic plus kinetic, and potential energy (bottom right). Most of the magnetic energy sits in a relatively small shell between $10^{-2}\,R_\odot < R_\mathrm{3D} < 10^{-1}\,R_\odot$. For the unequal-mass merger this is the region just outside the inert core of the initial primary white dwarf. At larger radii the magnetic pressure is almost comparable to the thermal pressure.}
    \label{fig:profile_remnant}
\end{figure*}

\begin{figure*}
    \centering
    \includegraphics[width=\textwidth]{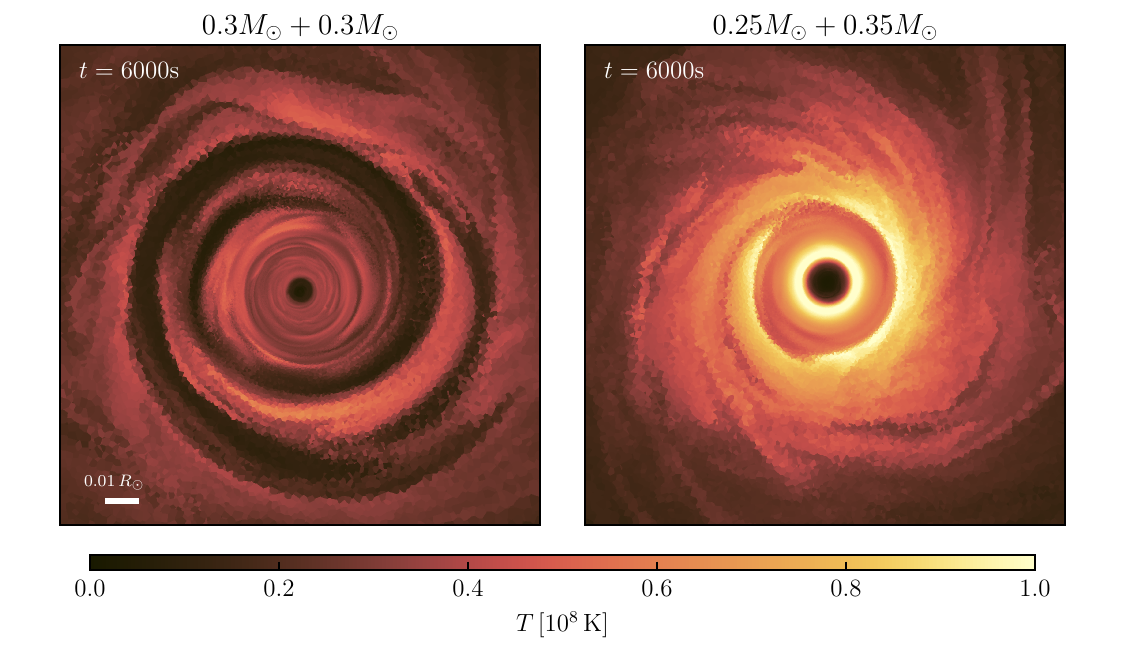}
    \caption{Temperature slices of merger remnants at the end of the simulation ($6000\,\mathrm{s}$) through the mid-plane. The equal-mass merger (left panel) is significantly colder and further away from ignition. We argue that it will likely eventually start helium burning in the centre, after compressing and heating up. The unequal-mass merger remnant (right panel) has already heated up a lot in a shell around its cold core, and will more likely ignite helium burning there first.}
    \label{fig:slices_temperature}
\end{figure*}

\begin{figure*}
    \centering
    \includegraphics[width=\textwidth]{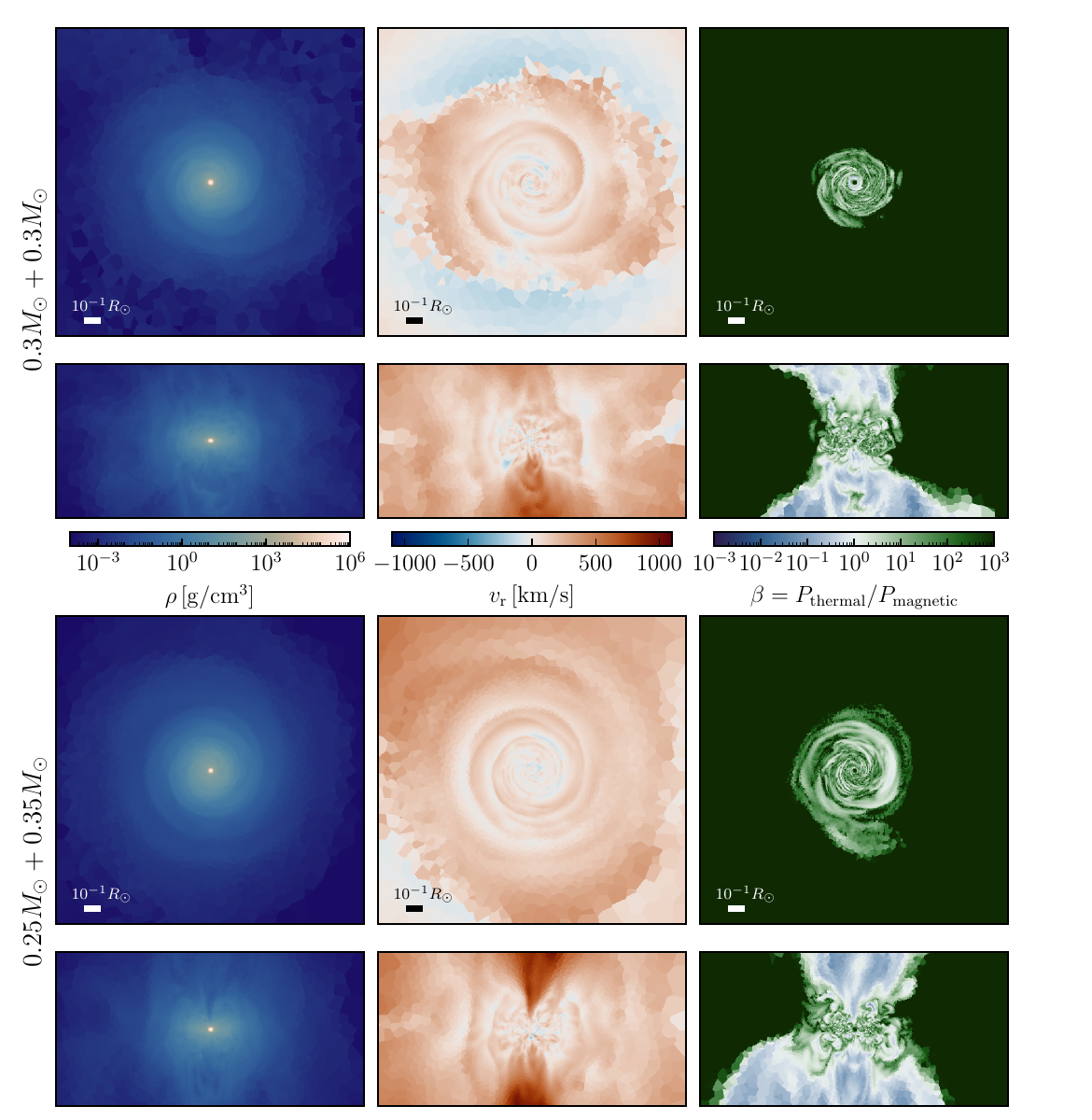}
    \caption{Slices showing a bipolar magnetically driven outflow on large scales for the remnants of the equal-mass merger (top half) and the unequal-mass merger (bottom half). The columns show density (left panels), radial velocity (middle panels), and plasma beta (right panels). Both remnants clearly generate strongly magnetised vertical outflows. Note, that they only emerge on large scales, at a height above $0.1\,R_\odot$.}
    \label{fig:outflow}
\end{figure*}

To qualitatively understand the evolution of the magnetic field in the merger remnants, we show slices of the azimuthal magnetic field strength on a linear scale at $2000\,\mathrm{s}$, $4000\,\mathrm{s}$, and $6000\,\mathrm{s}$ in Figure~\ref{fig:phi}. Both merger remnants start with an unordered small scale field with many field reversals at $2000\,\mathrm{s}$. This type of field is expected from dragging along the magnetic field previously amplified in a small-scale dynamo with the rotation. At $4000\,\mathrm{s}$ we see that most of the small-scale field reversals have disappeared and a large-scale ordered azimuthal field has emerged.  Moreover, the azimuthal field strength has increased. At $6000\,\mathrm{s}$ this evolution has continued for the equal-mass merger and the azimuthal field is coherently anti-aligned. In the unequal-mass merger, the structure of the azimuthal magnetic field at $6000\,\mathrm{s}$ is similar to its structure at $4000\,\mathrm{s}$, that is it still maintains a few field reversals on large scales. Moreover, its strength has decreased again.

We quantify the evolution of the magnetic field strength and in particular its azimuthal component for both merger remnants in Figure~\ref{fig:profile_bfld}. The left panel shows the profile of the total magnetic field strength in a cylinder with a height of $0.03\,R_\odot$, centred on the mid-plane. All our results are insensitive to the exact choice of the height of the cylinder. The magnetic field strength is computed as the volume-weighted average root mean square field. In other words, we compute the strength of a constant uniform magnetic field with the same total magnetic energy.

In the equal-mass merger the total magnetic field strength increases essentially monotonically at all radii with time. It saturates first at larger radii and last between $0.01\,R_\odot$ and $0.02\,R_\odot$, where it also reaches its highest strength. At $5000\,\mathrm{s}$, it has saturated at all radii and barely evolves anymore until the end of the simulation at $6000\,\mathrm{s}$.

The unequal-mass merger first shows a similar evolution of the magnetic field strength and reaches a profile similar to the final profile of the equal-mass merger already at $4000\,\mathrm{s}$. After that, however, its magnetic field strength declines again quickly at all radii. At the end of the simulation at $6000\,\mathrm{s}$, it has decreased to less than half of its peak strength.

To understand the connection between amplification and ordering of the azimuthal magnetic field, we look at the radial profiles of the energy density in the azimuthal component of the magnetic field in the middle and right columns of Figure~\ref{fig:profile_bfld}. We show the azimuthal magnetic energy density split by direction, that is the energy of the azimuthal magnetic field of all cells with a positive (top row) or negative (bottom row) value in each radial bin, divided by the total volume of each bin.

In the remnant of the equal-mass merger (middle column) the energy in the aligned and anti-aligned azimuthal magnetic field is initially comparable. The energy density in the aligned magnetic field then quickly decreases and the energy density in the anti-aligned magnetic field increases. At $4000\,\mathrm{s}$ the anti-aligned azimuthal field completely dominates over the aligned field and grows slightly further until it saturates at $5000\,\mathrm{s}$ and then remains constant. 

The remnant of the unequal-mass merger evolves similarly until $4000\,\mathrm{s}$. However, its then dominant anti-aligned azimuthal field decays again quickly afterwards and its energy density decreases  by more than a factor of four until $6000\,\mathrm{s}$. The anti-aligned azimuthal field continues to dominate over its aligned counterpart and the field remains ordered. 

The emergence of a dominant ordered azimuthal field in the equal-mass merger strongly points to the presence of a large-scale dynamo. We argue that it is driven by differential rotation ($\Omega$-effect) and the magnetorotational instability (MRI) ($\alpha$-effect), similar to recent results for proto-neutron stars \citep{ReboulSalze2022} and in the remnants of mergers of double neutron star binary systems for high resolution simulations \citep{Kenta2024}.

To directly show the presence of the MRI-driven large-scale dynamo in the stratified disk we show a butterfly diagram \citep{Brandenburg1995, Stone1996, Hirose2006, Gressel2010, Davis2010, Simon2011}, that has already been shown for idealised setups in \textsc{arepo} \citep{Pakmor2013b, Zier2022}. We show the butterfly diagram at a radius of $0.015\,\mathrm{R_\odot}$ in Figure~\ref{fig:butterfly} that shows outward moving patterns in the radial magnetic field that are characteristic for the MRI.

We follow \citet{Rembiasz2016} to estimate the fastest growing mode of the MRI at $2000\,\mathrm{s}$. We find $\lambda_\mathrm{mri}\,{\gtrsim}\,10^{-3}\,R_\odot$, which is well-resolved in both merger remnants with cells of $R_\mathrm{cell}\,{\sim}\, 10^{-4}\,R_\odot$. Similarly we find that the MRI should grow on a timescale of $t_\mathrm{mri}\,{\gtrsim}\,100\,\mathrm{s}$, in reasonable agreement to the growth rates we find in our simulation of several $100\,\mathrm{s}$.

Both merger remnants show the alternating patterns in the radial magnetic field moving outward from the mid-plane that are typical for the MRI. In the equal-mass merger these patterns emerge clearly around $2000\,\mathrm{s}$ and are accompanied by an ordering and amplification of the mean azimuthal magnetic field. They, as well as the strong azimuthal field, remain until the end of the simulation. Again, the unequal-mass merger shows a similar phenomenon between $2000\,\mathrm{s}$ and $4000\,\mathrm{s}$. After that, however, the MRI dies out and the azimuthal field becomes weaker again. The fast decay of the azimuthal magnetic field in the unequal-mass merger remnant at late times is likely a result of the numerical resistivity of our code and might be overestimated. In the equal-mass merger the ongoing MRI counteracts and prevents the same decay. Interestingly, the ordered azimuthal magnetic field is strongest slightly above and below the mid-plane in the equal-mass merger, but is strongest at the mid-plane in the unequal-mass merger.

To understand the quantitative differences between the evolution of both merger remnants, we look at the evolution of their angular velocity profiles in the left panel of Figure~\ref{fig:profile_omega}. At $2000\,\mathrm{s}$, they are very similar for $R_\mathrm{cyl}\,{\gtrsim}\,0.02\,R_\odot$ in both mergers. The inner part of the unequal-mass merger does not rotate, while the rotation profile continues smoothly deep into the core of the equal-mass merger. However, from there on the angular velocity profile of the equal-mass merger shows very little evolution. In contrast, the profile of the unequal-mass merger remnant evolves quickly and eventually reaches rigid rotation in most of the inner part of the remnant. Consistent with Figure~\ref{fig:butterfly}, the angular velocity profile at $0.015\,R_\odot$ has become flat at $5000\,\mathrm{s}$, and the MRI stops because its instability condition is not met anymore \citep{Balbus1991,Balbus1998}.

The total magnetic field, dominated by the azimuthal magnetic field, saturates with an energy density ${\sim}\,20\%$ of the rotational energy in the region that is unstable to the MRI and an order of magnitude stronger than the kinetic energy density in radial or vertical motions. In the centre, where the rotation profile is stable against the MRI, it saturates at a significantly lower value.

To better understand the evolution of the angular velocity profile we look at the evolution of the cumulative angular momentum profile in the right panel of Figure~\ref{fig:profile_omega}. Both merger remnants conserve total angular momentum well. They transport angular momentum to larger radii, far beyond the radial scales shown in Figure~\ref{fig:profile_omega}. The transport is completely dominated by the effective viscosity introduced from the magnetic fields (Maxwell stress), as verified by averaging the stresses over azimuthal shells at different times in the simulations. Transport via fluctuations of the velocity field (Reynolds stress) is essentially irrelevant. The equal-mass merger remnant contains and retains significantly more angular momentum in its inner region. Over longer timescales, however, it will likely similarly approach rigid body rotation.

\section{Discussion}
\label{sec:discussion}

Simulating the evolution of the merger remnant in full 3D with MHD is still only possible for a limited time, restricted by computing resources but also the accumulation of numerical errors (for example the numerical diffusion transporting magnetic fields and angular momentum into the core of the remnant). Nevertheless, we can now evolve the remnant long enough to discuss its potential future evolution, and in particular differences in the simulated and possible longer-term evolution of both merger remnants. Since the merger remnants have similar masses, we expect that they will eventually have similar convective cores by the time they settle down to the helium-core-burning main sequence \citep{Paczynski1971,Ostrowski2021}. However, the path to that structure may well be sufficiently different for the equal-mass and unequal-mass merger remnants as to lead to qualitative differences in their long-term magnetic properties.

We show profiles of both merger remnants at $6000\,\mathrm{s}$ in Figure~\ref{fig:profile_remnant}. Both remnants have very similar cumulative mass profiles. Most of their mass (${\sim}\,75\%$) is located within $0.02\,R_\odot$. The remaining $0.15\,M_\odot$ are spread out over a large volume out to a solar radius.

One fundamental difference is that the equal-mass merger is close to being fully mixed, only the innermost $10^{-2}\,M_\odot$ is strongly dominated by material of one of the white dwarfs. In contrast, the innermost $0.2\,M_\odot$ of the unequal-mass merger is the undisturbed and unmixed core of primary white dwarf. It is surrounded by $0.25\,M_\odot$ of mixed material. Finally, its outermost $0.15\,M_\odot$ are dominated by material originating from the secondary white dwarf.

Interestingly, most of the total angular momentum of both runs (${\sim}\,1.6{\times}10^{50}\,\mathrm{g\,cm^2\,s^{-1}}$) ends up in the very outermost layers of the merger remnants that contain little mass already in the initial dynamical merger. As shown in Figure~\ref{fig:profile_remnant} the equal-mass merger only has ${\sim}\,30\%$ and the unequal mass merger only ${\sim}\,15\%$ of their total angular momentum in the inner part of the remnant that contains ${\gtrsim}\,75\%$ of its total mass. How much of the angular momentum ends up in the central part of the remnant, however, is crucial for the evolution of its magnetic fields.

We show temperature slices through the mid-plane of both merger remnants at $6000\,\mathrm{s}$ in Figure~\ref{fig:slices_temperature}. The unequal-mass merger has a hot layer just outside its undisturbed, non-rotating (see Figure~\ref{fig:profile_omega}) cold core where material originating from both white dwarfs is mixed. We expect it to heat up further at this layer, by compression and potentially also dissipation of rotational energy, and eventually ignite helium fusion there. We will need detailed stellar evolution simulations that take into account the competition between electron thermal conduction and further heating in this shell to confirm or correct this picture. 

If helium burning ignites in this shell, it will generate a local convective shell there \citep{hewdsimon}. Importantly, most of the magnetic energy in the unequal-mass merger remnant is located in exactly this region. Nevertheless, the magnetic field is still dynamically completely irrelevant with a plasma beta of $\beta \sim 10^3$ (see lower right panel of Figure~\ref{fig:profile_remnant}). Therefore, we might assume that most of its ordered field in this region will be destroyed by the convective shell. Once the shell-burning recedes into the centre of the star, it will leave an unordered small-scale field behind by the time the star develops a steady-state helium-burning convective core.

Because the timescale on which the magnetic field dissipates via Ohmic resistivity scales with the square of the coherence scale of the magnetic field, the small-scale field left behind by the convective shell will decay orders of magnitudes faster than an ordered large-scale field. We can roughly estimate the timescale of Ohmic decay, assuming Spitzer conductivity, as 
\begin{equation}
    \tau \approx 10\,\mathrm{Myr}\left( \frac{R}{10^{-2}\,\mathrm{R_\odot}} \right)^2 \left( \frac{T}{10^6\,\mathrm{K}} \right)^{-3/2},
\end{equation}
where $R$ is the coherence scale of the magnetic field and $T$ the temperature. Thus, for a typical temperature above $10^7\,\mathrm{K}$ (see Figure~\ref{fig:slices_temperature}) a small-scale field will disappear quickly and after a million years the unequal-mass merger might not have any detectable trace of its former strong magnetic field left.

The equal-mass merger remnant has a significantly lower peak temperature at the end of the simulation. Moreover, its temperature is more homogeneous. Owing to its fully coupled, also differentially rotating core (see Figure~\ref{fig:profile_omega}), we may expect it to heat up in its centre most quickly and start helium fusion there. In this case it seems likely to us that the helium ignition will be so central as to directly lead to a convective helium-burning core, without the intermediate stage of inward-moving burning shells described above for the unequal-mass merger. For this post-merger mass, the convective core would probably cover its innermost $0.2\,M_\odot$ \citep{Paczynski1971,Ostrowski2021}. In this case, the region that contains most of its magnetic energy, that is dominated by a large-scale ordered azimuthal field, will be outside the convective core. Therefore, we might assume that its magnetic field could remain and -- because it is coherent on large scales -- it could still be seen many million years later.

We note that other magnetic dynamos that have been proposed to act in stars can in principle also be active in our merger remnants. In particular the Tayler-Spruit dynamo \citep{Spruit2002} that has recently seen a renewed focus in numerical simulations \citep{Braithwaite2006, Petitdemange2023a, Petitdemange2023b, Ji2023} might be present. However, it is outstripped by the faster MRI-driven large-scale dynamo that dominates the evolution of the magnetic field in our merger remnants.

The magnetic field strength in the main body of both merger remnants of $10^{10}\,\mathrm{G}$ (see Figure~\ref{fig:evolution_bfld}) is much higher than the surface field observed in the known magnetic sdO stars \citep[several $10^5\,\mathrm{G}$][]{Pelisoli2021}, but comparable to surface strengths of highly magnetised white dwarfs \citep{Bagnulo2022}. However, we cannot easily extrapolate from these values to observable surface field strengths. At radii larger than $\approx 0.03\,R_\odot$ the magnetic field is dynamically relevant in both merger remnants at the end of our simulations (see Figure~\ref{fig:profile_remnant}). If more angular momentum or magnetic energy are transported to the outer layers this will eventually cause the remnant to unbind its outer layers.

The main difference between both merger remnants stems from their mixing levels. The equal-mass merger is well mixed throughout, but the unequal-mass merger retains the undisturbed core of the primary white dwarf in its centre. Importantly, our result for the equal-mass merger is not limited to mergers of two white dwarfs with exactly the same mass. Rather, for a ratio of central densities of $\rho_\mathrm{c,secondary} / \rho_\mathrm{c,primary} > 0.6$ both white dwarfs are fully disrupted and mixed \citep{Zhu2013} and we expect them to behave similar to our simulation of the equal-mass merger.

The different rotation profiles that lead to the different evolutionary paths between equal and unequal-mass mergers are completely consistent with previous simulations of mergers of two carbon-oxygen white dwarfs at much lower resolution, that consistently showed a declining angular velocity profile for equal-mass mergers and a profile that is flat in the undisturbed core and has a broad peak in the envelope for unequal-mass mergers \citep{Zhu2013,Shiber2024}.

In principle, a similar dynamical evolution of the merger remnant as we see for our equal-mass merger would be expected for the merger of two carbon-oxygen white dwarfs presented in \citet{Zhu2015}. However, the simulation was not run long enough (it stopped before both cores fully merged, equivalent to the state of our equal-mass merger at roughly $1000\,\mathrm{s}$), and was also run at roughly $10$ times lower mass resolution than our simulations.

It is interesting to speculate that the main difference we see between mergers, that is a different rotation profile, might be a more general result. As discussed above it has already been seen for equal-mass mergers of two neutron stars \citep{Kenta2024}. It is potentially also relevant for mergers of main sequence stars \citep{Schneider2019}. If the rotation profile of the merger remnant is unstable to the MRI we would expect to see a similar large-scale dynamo in simulations with sufficient resolution that run long enough.

Some hot subdwarfs may also be formed via the merger of a helium white dwarf with a post-sdB `hybrid` white dwarf, that is a low-mass white dwarf with a carbon-oxygen core but a thick helium shell \citep{justhamco,Zenati2019,MillerBertolami+2022}. These mergers can also occur for equal-mass and unequal-mass white dwarf binaries, since hybrid white dwarfs can probably form in a significant range of masses. This could enable variations of the scenario we have investigated, possibly including equal-mass mergers which burn helium stably in a shell around an already helium-depleted carbon-oxygen core.

Both merger remnants generate bipolar highly magnetised outflows ($\beta{=}P_\mathrm{thermal}/P_\mathrm{magnetic}{<<}1$) at late times as shown in Figure~\ref{fig:outflow}. The outflows are qualitatively similar to those observed in other simulations of stellar mergers or common envelopes \citep{Ondratschek2022}, white dwarf--neutron star mergers \citep{MoranFraile2024}, or mergers of two neutron stars \citep{Kenta2024}. However, interestingly the outflows in our simulations are only launched at a distance of $\gtrsim 0.1\,R_\odot$, far outside the core of the merged object. The outflows shown in Figure~\ref{fig:outflow} originate from scales about ten times larger than the cores of the merger remnants shown in Figure~\ref{fig:slices_temperature}. The out-flowing material is less dense than its environment and magnetically dominated. Further study will be needed to  understand exactly which physical processes determine the properties of these outflows. The properties of outflows might in particular also be sensitive to the properties of the magnetic field prior to its amplification by the MRI \citep{Beckwith2008}.

Finally, it is important to briefly discuss the main limitations of our simulations. One obvious limitation is that we simulated only two different binary systems. We can extrapolate qualitatively if other binary systems with different mass ratios behave more like an equal-mass merger or an unequal-mass merger \citep{Zhu2013}. However, the evolution of the merger remnants might sensitively depend on the precise amount of angular momentum that ends up in the inner parts of the merged object (see Figure~\ref{fig:profile_omega}). Therefore, running a larger grid of models will be needed to confidently establish how stable our results are. We note that we ran our simulation in an almost identical setup three times for technical reasons, and obtained qualitatively similar results. The most interesting difference between reruns was that in one version the ordered azimuthal magnetic field in the unequal-mass merger remnant had a sign flip at the mid-plane. It is possible that the signs of the ordered magnetic field above and below the mid-plane decouple, but more work is needed to come to a firm conclusion.

It is interesting to compare our simulations to more idealised setups that only follow the evolution of magnetic fields after a merger of two white dwarfs has settled into an axisymmetric remnant, and assume axisymmetry \citep{Ji2013}. Compared to these simulations, we find magnetic fields of similar strength. They also find an active MRI, but without the large-scale dynamo that generates the dominant ordered azimuthal field in our simulations. They also find a bipolar outflow that is, however, launched much deeper into the remnant than in our simulation.

Other limitations are a result of the numerical schemes we employ. As discussed before, we do not include any explicit viscosity or resistivity. Therefore, both are dominated by their numerical counterpart that results from the finite volume scheme and operates on the scale of the mesh. The numerical viscosity and resistivity are therefore set by the numerical resolution we can afford, and are many orders of magnitude larger than their physical values. Therefore we overestimate dissipation effects like the decay of the magnetic field in the unequal-mass merger at late times (see Figure~\ref{fig:butterfly}).

Even more important is the ratio between viscosity and resistivity, the magnetic Prandtl number. In our simulations, the effective Prandtl number is close to unity, because numerical viscosity and resistivity both operate on the grid scale. Physical Prandtl numbers can deviate from unity by many orders of magnitude, though. In particular, flows with Prandtl numbers below unity might suppress or even prevent magnetic dynamos \citep{Schekochihin2004,Warnecke2023}. Assuming Spitzer resistivity \citep{Spitzer1953} and Braginskii viscosity \citep{Spitzer1962,Zuhone2015} we can crudely estimate the magnetic Prandtl number of a cell as
\begin{equation}
    \mathrm{Pr_m} \approx 10^{-28} \left( \frac{T}{\mathrm{K}} \right)^4 \left( \frac{\rho}{\mathrm{g\,cm^{-3}}} \right)^{-1}.
\end{equation}
There are two main regions that are interesting and important for magnetic field amplification in our simulations. The shear layer during the initial merger with a density $\rho\,<\,10^4\,\mathrm{g\,cm^{-3}}$ and temperature $T\,>\,10^7\,\mathrm{K}$, as well as the region where the MRI is most active, that has a typical density of  $\rho\,{\lesssim}\,10^5\,\mathrm{g\,cm^{-3}}$ and temperatures above $10^7\,\mathrm{K}$. Both have Prandtl numbers equal to or larger than unity. Therefore, the presence of an efficient magnetic dynamo in our simulations is expected. Moreover, even in regions with low Prandtl numbers the magnetic Reynolds number is still very large, so even there a dynamo may still operate \citep{Warnecke2023}.

Finally, numerical diffusion is probably the main reason for the transport of magnetic field lines and angular momentum into the unmixed cores of the unequal-mass merger.

\section{Summary and outlook}
\label{sec:summary}

We present two ideal MHD simulations of mergers of binary systems of two helium white dwarfs with the same total mass of $0.6\,M_\odot$. The simulations differ by the mass ratio of the initial white dwarfs: the equal-mass merger starts from a binary system of two helium white dwarfs of $0.3\,M_\odot$ each, and the unequal-mass merger from a binary system with a $0.25\,M_\odot$ and a $0.35\,M_\odot$ white dwarf.

We show that in both mergers the magnetic field is quickly amplified during the inspiral and the dynamical merger by a small-scale dynamo (Figure~\ref{fig:evolution}). The magnetic field strength saturates at similar field strengths of ${\sim}\,10^{10}\,\mathrm{G}$ in both mergers, and the merger remnants relax to the same density profile as shown in Figure~\ref{fig:profile_density}. Despite having the same total angular momentum, the equal-mass merger has twice the angular momentum in the inner parts of its remnant. The strong magnetic field helps transporting angular momentum outward and will slow down the new star.

We also show, however, that there are qualitative differences between the equal and unequal-mass mergers. As shown in Figure~\ref{fig:evolution} and in Figure~\ref{fig:profile_remnant}, the remnant of the equal-mass merger is mixed all the way to the centre. In contrast, the unequal-mass merger retains the inert non-rotating core of the primary white dwarf.

In both merger remnants, the MRI drives a large-scale dynamo (see Figure~\ref{fig:butterfly}), that creates an ordered azimuthal magnetic field that completely dominates after $50$ rotations. We find quantitative differences though, connected to the different rotation profiles in Figure~\ref{fig:profile_omega}. The equal-mass merger has and retains a declining rotation profile all through the object that allows the MRI to operate \citep{Balbus1991}. Outside its non-rotating core, the unequal-mass merger initially also has a declining rotation profile that drives an MRI and a large-scale dynamo generates an ordered azimuthal magnetic field. However, over $4000\,\mathrm{s}$ angular momentum transport transforms its rotation profile to essentially rigid rotation and the MRI dies out again. It is possible that a similar evolution will eventually happen for the equal-mass merger remnant as well, albeit on a slightly longer timescale.

We can, however, still only simulate the merger remnant dynamically for a very limited period of time. Therefore, the crucial next step is to follow up for many viscous timescales \citep{Schwab2018} and eventually much longer nuclear timescales \citep{Schneider2020} to understand the long-term evolution of the merger remnants. We nevertheless speculate that the remnants of the equal-mass merger and the unequal-mass merger might evolve in a fundamentally different way. We suggest that initial helium shell burning in the unequal-mass merger remnant will turn the ordered magnetic field into an unordered field that is then quickly dissipated. In contrast, the equal-mass merger remnant will likely directly ignite in the centre, and its strong, ordered magnetic field might survive long-term outside the convective core. Therefore, a large mass ratio of the merger might be the selecting factor that leads to highly magnetised sdO stars. More work is needed though to establish a firm conclusion if the strong magnetic fields in the merger remnants remain and if they are eventually observable on the surface of either star.

Finally, the strong magnetic fields we find in the outer parts of the envelope (see Figure~\ref{fig:profile_remnant}) and the magnetically driven bipolar outflow might directly lead a short, but luminous transient \citep{Beloborodov2014}.

\begin{acknowledgements}
We thank the referee for helpful comments that improved the paper. RP thanks Wilma Trick for help with the colour scheme of the paper. We also use colormaps from \citet{CrameriColormaps2021}. RP thanks Andrey Beloborodov, Thomas Guillet, Mike Lau, Valentin Skoutnev, Henk Spruit, and Volker Springel for helpful discussions. IP acknowledges support from a Royal Society University Research Fellowship (URF\textbackslash R1\textbackslash 231496). The work of FKR and FRNS is supported by the Klaus Tschira Foundation. FKR, MV, and RK acknowledge funding by the European Union (ERC, ExCEED, project number 101096243). Views and opinions expressed are however those of the authors only and do not necessarily reflect those of the European Union or the European Research Council Executive Agency. Neither the European Union nor the granting authority can be held responsible for them. This work has received funding from the European Research Council (ERC) under the European Union’s Horizon 2020 research and innovation programme (Grant agreement No.\ 945806). This work is supported by the Deutsche Forschungsgemeinschaft (DFG, German Research Foundation) under Germany’s Excellence Strategy EXC 2181/1-390900948 (the Heidelberg STRUCTURES Excellence Cluster).
\end{acknowledgements}

\bibliographystyle{aa}

\end{document}